\documentclass{article}

\usepackage{arxiv}

\usepackage[utf8]{inputenc} 
\usepackage[T1]{fontenc}    
\usepackage{hyperref}       
\usepackage{url}            
\usepackage{booktabs}       
\usepackage{amsfonts}       
\usepackage{nicefrac}       
\usepackage{microtype}      
\usepackage{lipsum}		
\usepackage{graphicx}
\usepackage{doi}
\usepackage{wrapfig}

\title{Decay rate measurements with a $^{137}$Cs radioisotope source at Jánossy Underground Research Laboratory (Csillebérc, Hungary)}

\author{ \href{https://orcid.org/0000-0003-2777-3719}{\includegraphics[scale=0.06]{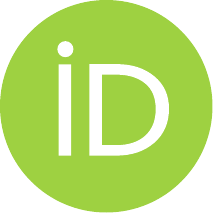}\hspace{1mm}Edit ~Fenyvesi}
    \\
	HUN-REN Wigner Research Centre for Physics\\
	Konkoly-Thege Miklós út 29-33., 1121 Budapest, Hungary \\
    HUN-REN Institute for Nuclear Research\\
	Bem tér 18/c, 4026 Debrecen, Hungary \\
	\texttt{fenyvesi.edit@wigner.hun-ren.hu} \\
	\And
	\href{https://orcid.org/0000-0001-9223-6480}{\includegraphics[scale=0.06]{orcid.pdf}\hspace{1mm}Gergely Gábor ~Barnaföldi} \\
	HUN-REN Wigner Research Centre for Physics\\
	Konkoly-Thege Miklós út 29-33., 1121 Budapest, Hungary \\
	\texttt{barnafoldi.gergely@wigner.hun-ren.hu} \\
	\AND
    \href{https://orcid.org/0000-0002-6872-916X}{\includegraphics[scale=0.06]{orcid.pdf}\hspace{1mm}Gábor Gyula ~Kiss} \\
	HUN-REN Institute for Nuclear Research\\
	Bem tér 18/c, 4026 Debrecen, Hungary \\
	\texttt{kiss.gabor@atomki.hun-ren.hu} \\
	\AND
    \href{https://orcid.org/0000-0000-0000-0000}{\includegraphics[scale=0.06]{orcid.pdf}\hspace{1mm}Dénes ~Molnár} \\
	HUN-REN Wigner Research Centre for Physics\\
	Konkoly-Thege Miklós út 29-33., 1121 Budapest, Hungary \\
	\texttt{molnar.denes@wigner.hun-ren.hu} \\
}

\renewcommand{\shorttitle}{Decay rate measurements $^{137}$Cs at Jánossy Underground Research Laboratory}

\hypersetup{
pdftitle={Decay rate measurements 137Cs at Jánossy Underground Research Laboratory},
pdfsubject={nucl-ex},
pdfauthor={Edit ~Fenyvesi, Gergely Gábor ~Barnaföldi, Gábor Gyula ~Kiss, Dénes ~Molnár},
pdfkeywords={decay rate measurements, 137Cs, Jánossy Underground Research Laboratory},
}

\begin{document}
\maketitle

\begin{abstract}
	The question whether an annual modulation is observable during nuclear decay rate measurements has long been the subject of research. One of the possible explanations for the annual variations would be the effect of solar neutrinos, the flux of which changes in correlation with the Earth-Sun distance. A decay rate measurement with a $^{137}$Cs source and a HPGe detector is currently being conducted 30 meters below the ground at Jánossy Underground Research Laboratory (Csillebérc, Hungary). The laboratory is part of the Vesztergombi High Energy Laboratory (VLAB), one of the TOP 50 research infrastructures in Hungary. From October 2022 to March 2023, data of six months' worth has been collected, and hence this is a new opportunity to check whether the annual variation in decay rate can be observed. The laboratory, the experiment, the data processing method, and the first results are presented in this study.
\end{abstract}

\keywords{Decay rate measurements \and $^{137}$Cs \and Jánossy Underground Research Laboratory}

\newpage

\section{Introduction}
Nuclear decays follows exponential law. High-precision detector techniques led us to investigate whether the decays of radionuclei are affected by weakly-coupled interactions. To deal with this, long-standing a measurement is required in a monitored and remote-controlled controlled underground laboratory to reduce the environmental and human effects.

In October 2022, measurements with a HPGe detector started at a low background counting site located 30~m below Earth's surface, at Jánossy Underground Research Laboratory (JURLAB) Csillebérc, Hungary). The main goal was to investigating the effect of environmental and seasonal changes on the background spectra. From October 2022 to March 2023, data of six
months’ worth has been collected, making it possible to study the deviation from the well-known exponential law of the radioactive decay involving $^{137}$Cs. 

Caesium is an optimal choice for the measurement, with a half life of 
$T_{1/2}= 30.018 \pm 0.08$~yr \cite{halflife2023}.
On one hand, $^{137}$Cs can decay in a spontaneous way: 
$^{137}$Cs $\rightarrow$ $^{137}$Ba + e$^{-}$ + $\bar{\nu}_{e}$.
On the other hand, electron neutrinos (mainly solar electron neutrinos) can interact with neutrons of the $^{137}$Cs nuclei: 
$\nu_{e}$ $+$ n$^{0}$ $\rightarrow$ e$^{-}$ $+$ b$^{+}$, 
and can decrease the number of $^{137}$Cs nuclei via the reaction 
$\nu_{e}$ + $^{137}$Cs $\rightarrow$ $^{137}$Ba $+$ e$^{-}$.
The number of $^{137}$Cs nuclei is decreased in the function of the time by spontaneous decay and the decay induced mainly by solar electron neutrinos.
The solar neutrino flux can vary according to the eccentricity of the orbit of Earth and the solar flair activity, so the following question arises: could any change of the half life of  $^{137}$Cs be observed via a counting experiment, or does the half life of $^{137}$Cs remain constant? 
Previously, mainly at low background counting sites, several long-term counting experiments have been started to study and interpret any deviation from the well-known exponential law of the radioactive decay involving various type of nuclei.
A recent paper 
\cite{Pomme2018decay} 
, citing several publications, presents a summary of the results obtained for $^{3}$H, $^{14}$C, $^{22}$Na, $^{54}$Mn, $^{65}$Zn, $^{90}$Sr, $^{109}$Cd, $^{134}$Cs, $^{152}$Eu, $^{209}$Cd, $^{226}$Ra and $^{241}$Am.
To date, no evidence of a deviation from the law of radioactive decay has been found so far.
In this work, the preliminary results of this study are presented.

\section {Measurements at Wigner Research Centre for Physics (Budapest, Hungary)}
\begin{wrapfigure}{r}{0.41\textwidth}
  \begin{center}
    \includegraphics[width=0.38\textwidth]{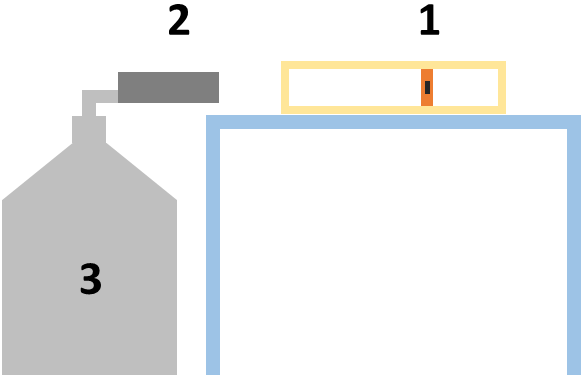}
  \end{center}
  \caption{Measurement setup at 30~m depth in the JURLAB. (See text for details.)}
  \label{setup}
\end{wrapfigure}
Jánossy Underground Research Laboratory is surrounded with Dachstein-type limestone.
40~cm thick outer walls made from concrete for nuclear reactors makes the site ideal for measurements that need low cosmic background. The experimental setup on Fig.~\ref{setup} contains:  
1) A $^{137}$Cs radiation source marked with a small black rectangle placed in a plastic case (orange rectangle). The case is placed in a  custom-made holder (yellow) allowing to change the distance between the sample and the detector. The distance was kept fixed during the measurement campaign presented here. 
2) high-purity Ge detector of relative efficiency of $\approx 40 \%$ at 1332~keV (model GC4018 from Canberra). 
3) The dewar containing the liquid nitrogen. 
The signal from the detector is collected by a LYNX-MCA \cite{Lynx2023} 
digital signal analyzer, which is not shown in the figure.
The analyzer is controlled remotely via its HTTP interface.

The gamma spectra were collected into histograms of 32768 bins. A linear energy calibration was used: $E=A+B \times k$, where $k$ is the $k$\textsuperscript{th} channel of the DAC ($0 \le k \le 32767$). Coefficients $A=0.068$~keV, $B= 0.177$~keV/channel, were calibrated to the $^{214}$Pb and $^{40}$K peaks at 351.9321~keV and 1460.802~keV, respectively, which are clearly visible in the background. The uncertainty of the linear energy calibration is about in the 10 eV order. Calibration was performed just once, in March 2022. Thereafter, $A$ and $B$ were kept fixed during the continuous measurement, i.e., no re-calibration was done midtime.

Background in a measured raw spectrum was estimated by fitting a quadratic polynomial of energy to the sidebands [650,657]~keV and [669, 674]~keV of the $^{137}$Cs peak.
After background subtraction, the area (total count) under the peak was determined using a Gaussian fit in the full [650,674]~keV range.
We did not have a dedicated setup for dead time measurement, so we relied on the live time estimates stored by the analyzer in the CNF output: $\Delta t_{\rm live} = \Delta t_{\rm nominal} - \Delta t_{\rm dead}$
Extracted background-subtracted peak areas were corrected for dead time by up-scaling to the nominal one-hour measurement duration by a factor $\Delta t_{\rm nominal} / \Delta t_{ \rm dead}$.

\section{Analysis}
\begin{figure}[h]
\centering
\begin{minipage}{.5\textwidth}
  \centering
  \includegraphics[width=.99\linewidth]{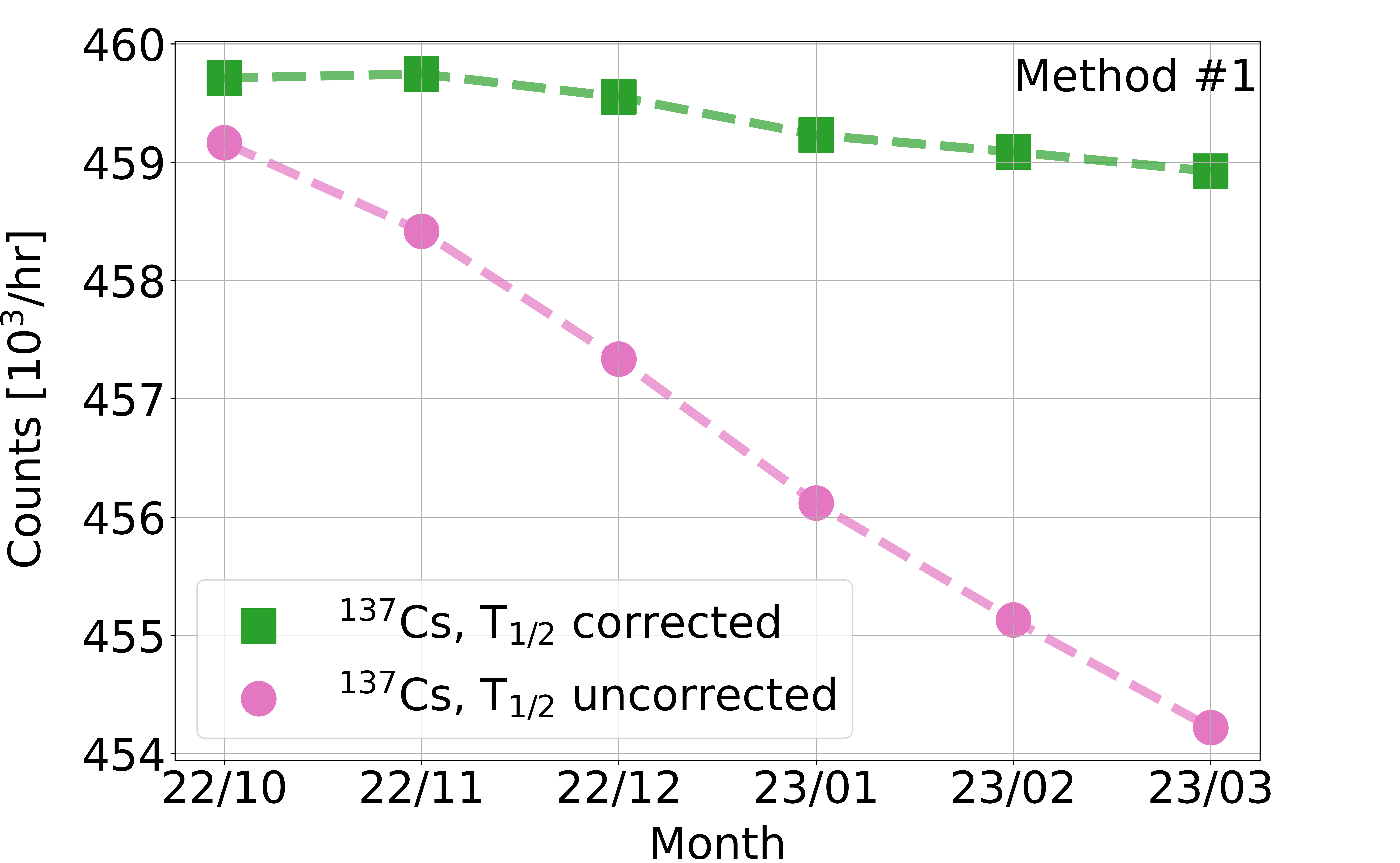}
  \caption{Results of analysis method No. 1.}
  \label{results1}
\end{minipage}%
\begin{minipage}{.5\textwidth}
  \centering
  \includegraphics[width=0.99\linewidth]{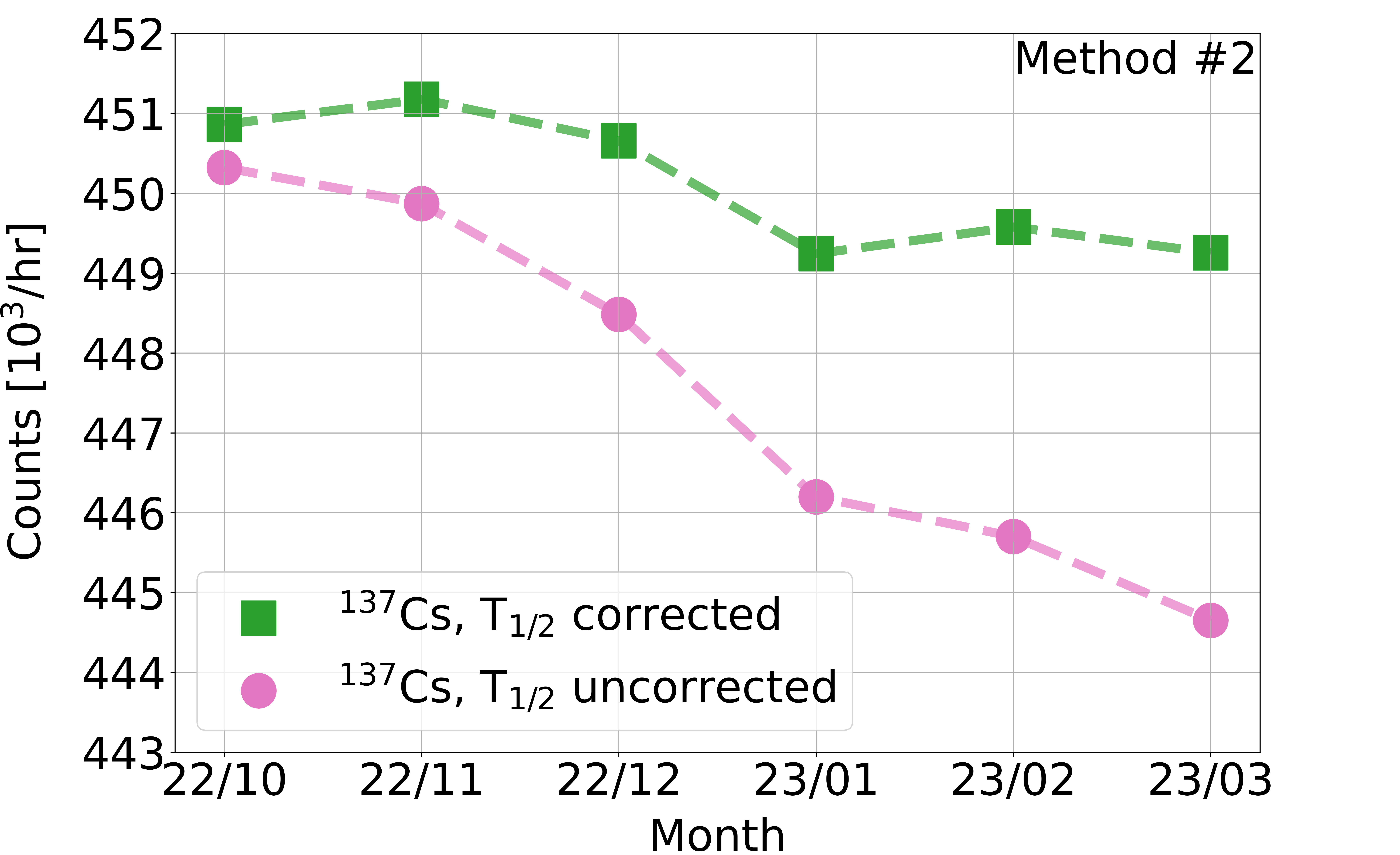}
  \caption{Results of analysis method No. 2.}
  \label{results2}
\end{minipage}
\end{figure}

According to the first method, each hourly dataset was analyzed separately.
Background-subtracted and dead-time corrected peak areas were tallied and then averaged as counts per hour for each month of the measurement: October 2022 to March 2023.
A gradual decreasing trend is manifest, as would be expected from a slow exponential decay of the source activity (Fig.~\ref{results1}).
Given the 30~yr half-life of $^{137}$Cs, source activity is expected to weaken annually by $\sim$ 1.1\%.
During a time interval $[t, t + \Delta t]$ the average source activity, relative to the activity at time $t_{0}$ is
\begin{equation}
    R\left(t, \Delta t|t_{0}\right) = \frac{1}{e^{-\lambda t_{0}}} \left[\frac{1}{\Delta t} \int_{t}^{t+\Delta t} dt' e^{-\lambda t'}\right] = e^{\lambda (t-t_{0})} \frac{1-e^{-\lambda \Delta t}}{\lambda \Delta t} \ \ ,
\end{equation}
where $\lambda \equiv \frac{\ln 2}{T_{1/2}}$.
The measured activity in each mini-run was scaled back, i.e., the extracted peak areas, to a common reference time $t_{0}$ by dividing by $R(t_{\rm nominal}, \Delta t|t_{0})$, where $t_{0}$ was started from 1\textsuperscript{st} Oct, 2022. The green graph on Fig.~\ref {results1} shows the result of this method. After correction, $\sim 0.2 \%$ decrease is still visible. This indicates that background subtraction and/or the assumption of Gaussian $^{137}$Cs peak vs energy are not accurate enough.

A second analysis method was tried as well: summing raw spectra for each month, and then doing background subtraction and area extraction on the monthly combined histograms.
Results can be seen on Fig. \ref{results2}.
If one does not compensate for the exponential weakening of the source (open circles), a clear decreasing trend is manifest that is similar in magnitude to the one of the first method --- albeit it does not look as smooth.
It is again straightforward to compensate for the source half-life, just slightly more complicated because histograms now contain monthly summed information.
One can use the ratio of the total uncompensated live time for the mini-run
$$t^{\rm TOT}_{\rm live} = \sum_{i} \Delta t_{\rm {live}, \it i}$$
where $i$ indexes the mini-runs, and the decay-compensated total live time
$$t^{\rm TOT, \rm comp}_{\rm live} = \sum_{i} R(t_{i}, \Delta t_{\rm nominal}|t_{0}) \Delta t_{\rm {live}, \it i}$$
to correct extracted areas by a multiplicative factor $t^{\rm TOT}_{\rm live}/ t^{\rm TOT, \rm comp}_{\rm live}$.
With the source half-life compensated for, the trend becomes flatter but a $\sim 0.4 \%$ decreasing residual still remains (Fig.~\ref{results2}).
Just like for Method 1, this suggests that background subtraction and/or the Gaussian shape assumption for the $^{137}$Cs peak are not accurate enough.


\section{Conclusion}
In this contribution we reported on on-going radioactive decay anomaly search measurement at the Jánossy Underground Research Laboratory of the Wigner Research Research Center at Csillebérc, Budapest, Hungary. We presented our preliminary analysis of gamma spectra collected more, than 6-month period for a $^{137}$Cs radioisotope. However a more detailed analysis is foreseen, by our analysis, data seems to be compatible with an annual modulation in $^{137}$Cs  decay rates that has relative amplitude of a few percent or more. We hope to significantly improve the sensitivity of the measurement after applying all the correction and post-calibration of the recorded data.

\section*{Acknowledgments}{Instrument R\&D and operation were done in the Jánossy Underground Research Laboratory (JURLAB) of the Vesztergombi Laboratory for High Energy Physics (VLAB) at Wigner RCP.}

\bibliographystyle{unsrtnat}






\end{document}